\documentclass{PoS}

\title{The Hard X-ray Spectral Evolution in X-ray Binaries, Active Galactic Nulcei and Ultraluminous X-ray Sources}

\ShortTitle{The Hard X-ray Spectral Evolution in XRBs, AGNs and
ULXs}
 
\author{Qingwen Wu\\ 
        International Center for Astrophysics, Korean
Astronomy and Space Science Institute, Daejeon 305348, Republic of
Korean\\
        E-mail: \email{qwwu@shao.ac.cn}}

\author{Minfeng Gu\\
Shanghai Astronomical Observatory, Chinese Academy of
Sciences, Shanghai, 200030, China\\
        E-mail: \email{gumf@shao.ac.cn}}

\abstract{ We explore the relationship between the hard X-ray photon
index $\Gamma$ and the Eddington ratio ($\xi=L_{X}(0.5-25\ \rm
keV)/L_{Edd}$) in six X-ray binaries (XRBs) with well-constrained
black hole masses and distances. We find that different XRBs follow
different anti-correlations between $\Gamma$ and $\xi$ when $\xi$ is
less than a critical value, while they follow the same positive
correlation when $\xi$ is larger than the critical value. This
anti-correlation and positive correlation are also found in low
luminosity active galactic nuclei (LLAGNs) and luminous QSOs
respectively. The anti-correlation and positive correlation of
different XRBs roughly converge to the same point ($\log
\xi=-2.1\pm0.2, \Gamma=1.5\pm 0.1$), which may correspond to the
accretion mode transition, since that the anti-correlation and
positive correlation between $\Gamma$ and $\xi$ are consistent with
the prediction of advection dominated accretion flows (ADAFs) and
standard disk/corona system respectively. We note that traditional
low-hard state are divided into two parts by the cross point $\log
\xi\sim-2.1$, i.e., \emph{faint-hard state} in the anti-correlation
part and \emph{bright-hard state} in the positive correlation part.
The accretion process in the \emph{bright-hard state} may be still
the \emph{standard accretion disk} as that in the high/soft state,
which is consistent with that both the cold disk component and broad
Fe K emission line are observed in some bright-hard state of XRBs
(e.g., GX 339-4, Miller et al. 2006). The \emph{ADAF} is only
important in the \emph{faint-hard state} XRBs.

 Motivated by the similarities of the state transition and timing
  properties of the ultraluminous X-ray sources (ULXs) to that of XRBs, we then
 constrain the black hole masses for seven
 luminous ULXs assuming that their X-ray spectral evolution is similar
  to that of XRBs (i.e., photon index is only depend on the Eddington ratio).
  We find that the BH masses of these seven ULXs are around
  $10^{4}M_{\odot}$, which are typical intermediate-mass BHs (IMBHs).
  Our results are roughly consistent with the BH masses constrained from
  the model fitting with a multi-color disk and/or the timing
  properties (e.g., QPO and break frequency).

    }

\FullConference{VII Microquasar Workshop: Microquasars and Beyond\\
         September 1-5 2008\\
         Foca, Izmir, Turkey}

\begin{document}

\section{Introduction}

   It is generally assumed that black hole (BH) X-ray
binaries (XRBs), active galactic nuclei (AGNs) and other BH systems
have a similar central engine, namely the central BH, the accretion
flow and a relativistic jet (for reviews see Fender et al. 2007, and
$\rm K\ddot{o}rding$ et al. in this proceeding). The similarities
are supported by recent two ``fundamental planes" of the BH activity
on all mass scales, which include $L_{\rm radio}-L_{\rm Xray}-M_{\rm
BH}$ (e.g., Merloni et al. 2003; Falcke et al. 2004) and $L_{\rm
bol}-M_{\rm BH}-T_{\rm break}$ (McHardy et al. 2006). The
hardness-intensity diagram and pattern of radio loudness in XRBs and
AGNs also support this unification scenario (e.g., $\rm
K\ddot{o}rding$ et al. 2006).

    The low/hard state and high/soft state are two main states in XRBs. Typically, the X-ray spectrum in
   hard state can be described as a power law with a photon index
   $\Gamma=1.5-2.1$, while the energy spectrum can be
   described with a thermal disk component and a power law with a photon index
   $\Gamma=2.1-4.8$ in soft state (e.g., Remillard \& McClintock
   2006). The hard X-ray photon index is anti-correlated to the Eddington ratio in some low/hard
   state XRBs (e.g., Yamaoka et al. 2005) and a sample of low
   luminosity AGNs (LLAGNs, Gu \& Cao 2008) when the Eddington ratio
   is less than a critical value. However, a positive correlation is found in XRBs with
    higher Eddington ratio (e.g., Kubota \& Makishima 2004)
   and luminous QSOs (e.g., Shemmer et al. 2006).

   Ultraluminous X-ray sources (ULXs) are pointlike, nonnuclear X-ray
  sources with X-ray luminosities between $10^{39}$ and $10^{41}\rm ergs/s$, well in excess of
  the Eddington limit for a stellar mass black hole.
  The true nature of ULXs is still unclear, especially their BH
  mass, and current models include two main alternatives: (1) ``intermediate-mass BHs" (IMBHs) with
  mass $M_{\rm BH}\simeq 10^{2}-10^{5} M_{\odot}$ (e.g., Colbert et al. 1999);
  (2) stellar mass BHs (e.g., Mirabel \& Rodriguez 1999).

\section{Sample}

   Six XRBs observed by $Rossi\ Xray\ Timing\ Explorer\ (RXTE)$ are selected,
    which have well determined hard X-ray spectral slopes, BH masses and distances.
    The X-ray data of all XRBs were observed during the \emph{decay} of the
    outburst, i.e., transition from soft to hard state. For comparison, 55 LLAGNs (27 LINERs + 28
    Seyferts), 26 radio-quiet QSOs, and 8 luminous ULXs are also selected.  These luminous ULXs have
     the 0.3-10 keV X-ray luminosities near or larger than $10^{40}\rm erg/s$, and have
     multiple (at least three epochs) high resolution $XMM-Newton$ and/or $Chandra$
      observations (see Wu \& Gu 2008 for more details and references therein).

\section{Results and Discussion}
\subsection{The hard X-ray spectral evolution in XRBs and AGNs}
  We present the relation between the photon
  index, $\Gamma$, and the Eddington ratio, $\xi$, of the XRB sample in Figure 1a,
   where the photon index $\Gamma$ is fitted in the 3-25 keV band.
   We define the Eddington ratio $\xi=L_{X}(0.5-25\ \rm keV)/\it L_{\rm Edd}$, in which $L_{X}(0.5-25\ \rm keV)$
  is extrapolated from the unabsorbed 3-25 keV band luminosity, since which was regarded
   as more represent the bolometric luminosity. It can be clearly seen that
  the X-ray photon index $\Gamma$ is strongly correlated with the
  Eddington ratio $\xi$ when $-2\lesssim\log \xi\lesssim0$. In the low Eddington ratio (e.g., $\log \xi\lesssim
  -2$) part, the anti-correlation is generally present, which actually has already been found in
  several occasions (e.g., Yamaoka et al. 2005; Yuan et al. 2007). However, the
  most remarkable result in our work is that different sources follow
  the different anti-correlations (Fig. 1a). We fit the anti-correlation points with linear least-square
  method for each source. We also fit all data points at the positive correlation ($-2\lesssim \log \xi \lesssim
  -1$) region as a whole with the same method (short dashed line in Fig.
  1a). Therefore, for each source, we obtained the cross point
  between the positive and anti-correlation part.
  Although the anti-correlation varies from source to source, we
  find the cross points of all XRBs roughly converge to the same point with small scatter
  ($\log \xi=-2.1\pm0.2, \Gamma=1.5\pm 0.1$).

\begin{figure}
\includegraphics[width=.99\textwidth]{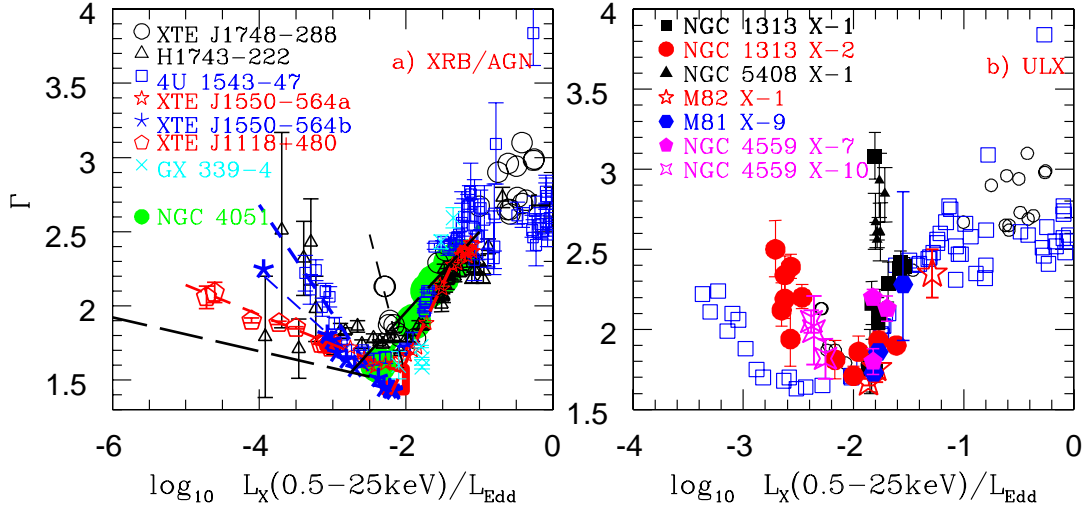}
\caption{(a) The relation between the X-ray photon index and the
Eddington ratio in XRBs and AGNs. The short dashed lines are the
linear least-square fits of XRBs,  and long-dashed lines are for
LLAGNs (left) and QSOs (right) respectively. (b) The relation
between the X-ray photon index and the Eddington ratio in ULXs. For
comparison, two XRBs (XTE J1748-288 and 4U 1543-47) are also
plotted.} \label{fig1}
\end{figure}

  For comparison, we also plot the best
  fits to the relation between the photon index and the Eddington
  ratio in LLAGNs (long dashed-line in Fig. 1a, Gu \& Cao 2008) and
  luminous QSOs (long dashed-line in Fig. 1a, Shemmer et al. 2006),
   We also include the Seyfert galaxy NGC 4051, which show strong
  X-ray spectrum and flux variation (green solid circles in Fig.
  1a). We find that the photon index is anti-correlated to the
  Eddington ratio in LLAGNs, while a positive correlation is present
  in QSOs and NGC 4051.
  The anti-correlation in LLAGNs and
  the positive correlation in luminous AGNs cross at point ($\log \xi=-2.78, \Gamma=1.61$), where the Eddington
  ratio $\xi$ is several times less than that of XRBs, which is mainly
   caused by the different bolometric correction factor (see Wu \& Gu 2008 for details).
  Therefore, it seems that the hard X-ray spectral
  evolution are also similar in XRBs and AGNs, which can be formulized as:
  \begin{equation}
  \Gamma = \kappa (\xi+2.1^{+0.7}_{-0.2}) +1.5^{+0.1}_{-0.1}
  \end{equation}
where $\kappa$ is the slope and $\xi$ is the Eddington ratio.

\subsection{The possible accretion process and the so-called ``hard
state problem" }
 Both the UV/optical bumps observed in QSOs and the soft X-rays
observed in high state XRBs can be naturally interpreted as
blackbody emission from the standard accretion disk (Shakura \&
Sunyaev 1976). ADAF is a hot, optically thin, geometrically thick
accretion flow, which can successfully explain most features of the
LLAGNs and low/hard-state XRBs (see Ho 2008; Narayan \& McClintock
2008 for recent reviews). The qualitative behavior of the
\emph{anti-correlation} between the photon index and Eddington ratio
is consistent with the predictions of the \emph{ADAF} spectrum
(e.g., Esin et al. 1997). The Comptonization of thermal synchrotron
photons in the ADAF is the dominated cooling mechanism at low
Eddington ratios. As the accretion rate increases, the optical depth
of the ADAF also increases, increasing the Compton $y$-parameter,
thereby leading to a harder X-ray spectrum. However, the
\emph{positive correlation} is consistent with the \emph{standard
disk-corona} model (e.g., Janiuk \& Czerny 2000). As the accretion
rate increases, both the fraction of accreting energy released to
the corona and electron temperature decreases, the corona becomes
weak and the \emph{optical depth} decreases, reducing the
y-parameter, and leading to a softer X-ray spectrum.

Recently, the observations of the cool disk component and the
relativistic broad Fe K emission line in hard state of several XRBs
(e.g., GX 339-4, Miller et al. 2006; Tomisic et al. 2008) have
suggested that the standard disk may extend to the innermost stable
circular orbit, which challenged the popular ADAF model in the hard
state (so-called ``\emph{hard state problems}", see a review of
Tomsick et al. in this proceeding). It is interesting to note that
the traditional hard state are divided into two parts, i.e.,
\emph{bright hard state} and \emph{faint hard state}, by the cross
point($\log \xi \sim -2.1\pm0.2$) of the anti-correlation and the
positive correlation. The X-ray spectral evolution in the bright
hard state is the same as that of high/soft state of XRBs and
luminous QSOs. It implies that accretion process in this
\emph{bright-hard state} may be the \emph{standard accretion disk}.
It is consistent with the cool disk component and the relativistic
broad Fe K emission line in the several bright hard state of XRBs
(e.g., Miller et al. 2006). In contrast, the \emph{ADAF model}
becomes important in the \emph{faint-hard state} of the XRBs with
$\log \xi \lesssim -2.1$. If this is the case, the cross point may
correspond to the accretion mode transition, which, however, is
different from the popular idea that the disk transition occurs at
$\Gamma=2.1$.

\subsection{The X-ray spectral evolution in ULXs and constraints on
their BH masses}
   It is found that the X-ray photon index of NGC 1313 X-1 is
    positively correlated to the luminosity, while these two parameters are anti-correlated in NGC 1313
    X-2  (Feng \& Kaaret 2006), which are similar to that of XRBs/AGNs.
    This phenomenon has also been found in many other
    ULXs (e.g., Roberts et al. 2004). The similarity between XRBs, ULXs and AGNs implies that
    there may be similar physics behind the phenomena, which motivates us to explore
    the properties of ULXs utilizing the unification of ULXs and XRBs (and AGNs).
    Specifically, the BH masses of ULXs can be constrained if we
    assume that the BH accretion is scale-free and that their X-ray spectral evolution
     is only determined by the Eddington ratio. Assuming the spectral evolution can be described by Eq. 3.1,
    we thus calculate the BH masses of ULXs through the
    least-squares linear fit on the available data for each ULX.
    Our results show that all the BHs in these luminous ULXs are IMBHs of around $(4-30)\times 10^{3}M_{\odot}$ (Table
    1). We find that our estimated BH masses are consistent within a
    factor of two with that derived either using $\nu_{\rm break}-M_{\rm BH}-L_{ \rm Bol}$ relation
   for XRBs and AGNs (McHardy et al. 2006) or from the multi-color disk fittings (Miller et al.
   2004) (see Table 1).  These consistency may support the validity of our
   method, which further implies that the assumption of similarity
  between ULXs and XRBs is likely reasonable.

\begin{table}
\caption{BH mass of ULXs.} \label{tab1}
\begin{tabular}{lcc|lcc}
\hline \hline
Source Name & Our BH mass & other BH mass$^1$ & Source Name & Our BH mass & other BH mass$^2$\\
\hline
 M81 X-9      &  $6.6\times10^{3}$   &  $5^{+7}_{-2}\times10^{3}$   &  NGC 5408 X-1  &  $<4.0\times10^{3}$  &  $2.5\pm1\times10^{3}$   \\
 M82 X-1      &  $6.6\times10^{3}$   &  $5^{+7}_{-2}\times10^{3}$   &  NGC 4559 X-7  &  $1.4\times10^{4}$   &  $7.6\times10^{3}$     \\
 NGC 1313 X-1 &  $7.9\times10^{3}$   &  $4^{+2}_{-1}\times10^{3}$   &  NGC 4559 X-10 &  $2.7\times10^{4}$   & ...    \\
 NGC 1313 X-2 &  $2.4\times10^{4}$   &  ...                  \\
\hline
\end{tabular}
\footnotesize (1)BH mass derived from the multicolor disk fittings;
(2) BH mass derived from the timing properties in this work.
\end{table}

\section*{Acknowledgments} Q.W.W. appreciate the LOC for the partial financial support.
This work is supported by National Science Foundation of China
(grants 10633010, 10703009 and 10833002) and by 973 Program (No.
2009CB824800), and a postdoctoral fellowship of the KASI.

\end{document}